\documentclass[12pt]{iopart}
\usepackage{iopams}
\usepackage{amsfonts}
\usepackage{graphicx}
\usepackage{subfigure}
\usepackage{amsthm}
\usepackage{times}
\usepackage{color}
\usepackage{ulem}
\usepackage[compress]{cite}

\newcommand{\eqref}[1]{{(\ref{#1})}}
\begin{document}


\title[]{Dark-like states for the multi-qubit and multi-photon Rabi models}

\author{Jie Peng$^{1}$, Chenxiong zheng$^{1}$, Guangjie Guo$^{2}$, Xiaoyong Guo$^{3}$, Xin Zhang $^{4,5}$, Chaosheng Deng$^{1}$, Guoxing Ju$^{4}$, Zhongzhou Ren$^{4,5,6,7}$, Lucas Lamata$^{8}$, Enrique Solano$^{8,9}$}
\address{$^{1}$Laboratory for Quantum Engineering and Micro-Nano Energy Technology and School of Physics and Optoelectronics, Xiangtan University, Hunan 411105, China}
\address{$^{2}$Department of Physics, Xingtai University, Xingtai 054001, China}
\address{$^{3}$School of Science, Tianjin University of Science and Technology, Tianjin 300457, China}
\address{$^{4}$Key Laboratory of Modern Acoustics and Department of Physics, Nanjing University, Nanjing 210093, China}
\address{$^{5}$Joint Center of Nuclear Science and Technology, Nanjing University, Nanjing 210093, China}
\address{$^{6}$Center of Theoretical Nuclear Physics, National  Laboratory of Heavy-Ion Accelerator, Lanzhou 730000, China}
\address{$^{7}$Kavli Institute for Theoretical Physics China, Beijing 100190, China}
\address{$^{8}$Department of Physical Chemistry, University of the Basque Country
UPV/EHU, Apartado 644, E-48080 Bilbao, Spain}
\address{$^{9}$ IKERBASQUE, Basque Foundation for Science, Maria Diaz de Haro 3, 48013 Bilbao, Spain}

\ead{\mailto{jpeng@xtu.edu.cn},\mailto{jugx@nju.edu.cn},\mailto{zren@nju.edu.cn}}
\date{\today}

\begin{abstract}
There are well-known ¡°dark states¡± in the even-qubit Dicke models,
which are the products of the two-qubit singlets and a Fock state,
where the qubits are decoupled from the photon field. These spin
singlets can be used to store quantum correlations since they
preserve entanglement even under dissipation, driving and
dipole-dipole interactions. One of the features for these ¡°dark
states¡± is that their eigenenergies are independent of the
qubit-photon coupling strength. We have obtained a novel kind of
dark-like states for the multi-qubit and multi-photon Rabi models,
whose eigenenergies are also constant in the whole coupling regime.
Unlike the ¡°dark states¡±, the qubits and photon field are coupled
in the dark-like states. Furthermore, the photon numbers are bounded
from above commonly at $1$, which is different from that for the
one-qubit case. The existence conditions of the dark-like states are
simpler than exact isolated solutions, and may be fine tuned in
experiments. While the single-qubit and multi-photon Rabi model is
well-defined only if the photon number $M\leq2$ and the coupling
strength is below a certain critical value, the dark-like
eigenstates for multi-qubit and multi-photon Rabi model still exist,
regardless of these constraints. In view of these properties of the
dark-like states, they may find similar applications like ``dark
states'' in quantum information.
\end{abstract}


\maketitle

\section{Introduction}\label{intro}
\noindent The Rabi model \cite{rabi} has been born for 80 years
\cite{80}. With semi-classical \cite{rabi} and fully quantized
versions \cite{jc}, it has found wide applications in magnetic
resonance \cite{rabi}, solid state \cite{irish}, quantum optics
\cite{guo}, cavity QED \cite{trapped}, circuit QED \cite{walraf} and
quantum information \cite{blais}. Although the quantum Rabi model
has a very simple form, describing the simplest interaction between
light and matter, its analytical solution had not been found until
2011 by Braak \cite{br}. This is partly due to the fact that there
is no closed subspace in its Fock space, which is different from
that in the Jaynes-Cummings model \cite{jc} with the rotating wave
approximation \cite{yo}. The qubit-photon ultrastrong coupling
regime has been reached in recent circuit QED experiment \cite{nie}.
However, in this regime, the Jaynes-Cummings model fails, so many
researches then focus on the Rabi model, which include the
analytical solution of the Rabi model retrieved by Chen et al using
Bogoliubov operators \cite{qing}, two-photon \cite{trave,chen,pj1},
two-qubit \cite{pj,2bite,pj2,qinghuchen,duan} and multi-qubit
\cite{dicke,heshu,zhangyunbo} generalizations, exact real time
dynamics \cite{bra1,bra2a}, deep strong coupling \cite{deep},
anisotropic Rabi model \cite{xie} and so on
\cite{zhong,murry,ultra}.

For the single qubit Rabi model, the eigenstates consist of infinite
photon number states, because there are no closed subspaces in the
Fock space \cite{br}. But this is not the case for the multi-qubit
case, because more qubits will bring closed subspaces. For example,
Rodr\'iguez-Lara et al has found ``trapping states'' (``dark
states'') \cite{rod} in the even qubit Dicke model, where two
identical qubits form a spin singlet and the eigenstates are just
products of these singlets and a Fock state. These singlets are
decoupled from the photon field, and will survive even under
dissipation, driving and also dipole-dipole interactions. So they
can be used to store quantum correlations. Since the qubits and
photon are decoupled, the eigenenergies of the ``dark states'' are
constants in the whole qubit-photon coupling regime, which
correspond to horizontal lines in the spectrum.

There has been some researches on the ``dark-like'' states of the
two-qubit and single photon Rabi model \cite{pj2,alge}. In this
paper, we will study multi-qubit and multi-photon Rabi model, and
show that ``dark-like'' eigenstates commonly exist, surprisingly not
just for even-qubit, but also odd-qubit, and multi-photon cases. The
single-qubit and multi-photon Rabi model is well-defined only if the photon
number $M\leq2$ and the coupling strength is below a certain critical value \cite{tp1},
but with multi-qubit it will bring about closed subspaces and the dark-like eigenstates
still exist in the whole coupling regime and for $M>2$. These dark-like states posses several features. Firstly, they exist
in the whole coupling regime with constant eigenenergies, just like
the ``dark states''. But surprisingly, the qubit and photon are not
decoupled and the wavefunctions are coupling dependent. Secondly,
the photon numbers in the eigenstates are bounded from above at $K$.
In particular, $K=1$ for the single-photon case. Thirdly, their
existence conditions are simpler than exact isolated solutions,
because they can be realized in arbitrary coupling regime with the
same qubit energy, which may be fine tuned in experiment. So just
like the ``dark states'', these ``dark-like'' states may get
possible application in quantum information.

The paper is organized as follows. In section \ref{s2}, we search
for the ``dark-like'' eigenstates for the multi-qubit Rabi model. In
section \ref{s3}, we generalize our study to the multi-qubit and
multi-photon Rabi models. In section \ref{SecImplementation} we give some experimental considerations for the implementation in quantum controllable platforms. Finally, we give our conclusions in
section \ref{s4}.

\section{Dark-like states for the multi-qubit Rabi model}\label{s2}
The Hamiltonian of the N-qubit quantum Rabi model reads ($\hbar=1$)
\cite{2bite,pj2}
\begin{equation}\label{gq}
H_{NQ}=\omega
a^{\dagger}a+\sum_{i=1}^{N}g_{i}\sigma_{ix}(a+a^{\dagger})
+\sum_{i=1}^{N}\Delta_i\sigma_{iz},
\end{equation}
where $a^{\dagger}$ and $a$ are the single mode photon creation and
annihilation operators with frequency $\omega$, respectively.
$\sigma_{i \alpha}, (\alpha=x,y,z)$ are the Pauli matrices
corresponding to the $i$-th qubit. $2\Delta_i$ is the energy level
splitting of the $i$-th qubit, and $g_{i}$ is the qubit-photon
coupling constant. $\omega$ is set to $1$ in the following
discussion.

The Hamiltonian (\ref{gq}) is usually infinite dimensional in the
Fock space, which is exactly the case for just one qubit, but with
more qubits it will bring about possible closed subspace. Working on
this finite dimensional subspace, we can obtain the solution of the
Hamiltonian (\ref{gq}) with finite photon numbers and the dark-like
eigenstates. For this purpose, we must first search for the
existence condition of this closed subspace. $H_{NQ}$ possesses a
$\mathbb{Z}_2$ symmetry with the transformation $R=\exp(i\pi
a^\dagger a)\prod_{i=1}^{N}\sigma_{iz}$, so we have
\begin{equation}
R|p\rangle=p|p\rangle
\end{equation}
with $p=\pm 1$.
At the same time, we can categorize the N-qubit states
$\{|\psi\rangle_{Nq}\}$ into two sets with the eigenvuales of
$\prod_{i=1}^{N}\sigma_{iz}$ being $1$ and $-1$ respectively, and
they are denoted by $2^{N-1}$ dimensional row vectors
$(|\psi\rangle_{Nq+})$ and $(|\psi\rangle_{Nq-})$ . It is easy to
find the following relations
\begin{eqnarray}
(|\psi\rangle_{Nq+})=(|\psi\rangle_{N-1~q-}\otimes|\downarrow\rangle_N,|\psi\rangle_{N-1~q+}\otimes|\uparrow\rangle_N),\label{nq1}\\
(|\psi\rangle_{Nq-})=(|\psi\rangle_{N-1~q+}\otimes|\downarrow\rangle_N,|\psi\rangle_{N-1~q-}\otimes|\uparrow\rangle_N),\label{nq2}
\end{eqnarray}
with the initial states $|\psi\rangle_{1q+}=|\uparrow\rangle_1$ and
$|\psi\rangle_{1q-}=|\downarrow\rangle_1$. Then we have two
unconnected subspaces
\begin{eqnarray}
|0, \psi_{Nq+}\rangle\leftrightarrow |1,
\psi_{Nq-}\rangle\leftrightarrow |2,
\psi_{Nq+}\rangle\leftrightarrow \cdots ~(p=1)\\
|0, \psi_{Nq-}\rangle\leftrightarrow |1,
\psi_{Nq+}\rangle\leftrightarrow |2,
\psi_{Nq-}\rangle\leftrightarrow \cdots ~(p=-1)
\end{eqnarray}
Only neighboring states within each parity chain are connected, so
$H_{NQ}^{\pm}$ will take the following form in even ($p=+1$) or odd
($p=-1$) subspace
\begin{eqnarray}\label{off}
H_{NQ}^\pm=\left(
  \begin{array}{cccccc}
   D_{N0}^\pm &O_{N0}^\pm &0 & 0&0&\dots\\
    O_{N0}^\pm & D_{N1}^\pm&O_{N1}^\pm & 0 &0&\dots  \\
   0&O_{N1}^\pm &D_{N2}^\pm & O_{N2}^\pm &0&\dots \\
     \dots&\dots&\dots&\dots&\dots&\dots\\
  \end{array}
\right),
\end{eqnarray}
where $D_{Nj}^\pm$ and $O_{Nj}^\pm$ ($j=0,1,2,3,\ldots$) can be
written as
\begin{eqnarray}
D_{Nj}^\pm=(\langle
j,\psi_{Nq\pm(-1)^j}|)^T H_{NQ}(|j,\psi_{Nq\pm(-1)^j}\rangle),\label{d}\\
O_{Nj}^\pm=(\langle j+1,\psi_{Nq\mp(-1)^j}|)^T
H_{NQ}(|j,\psi_{Nq\pm(-1)^j}\rangle),\label{o}
\end{eqnarray}
where $(|j,\psi_{Nq\pm(-1)^j}\rangle)$ is a $2^{N-1}$ dimensional
vector, and $D_{Nj}^\pm$, $O_{Nj}^\pm$ are $2^{N-1}\times 2^{N-1}$
matrixes. Substituting Eqs. \eqref{nq1} and \eqref{nq2} into Eqs.
\eqref{d} and \eqref{o}, we get the following expressions for
$D_{Nj}^\pm$ and $O_{Nj}^\pm$,
\begin{eqnarray}
D_{Nj}^\pm=&(\langle j,\psi_{N-1~q\mp(-1)^j}|\otimes{}_N\langle
\downarrow|,\langle
j,\psi_{N-1~q\pm(-1)^j}|\otimes{}_N\langle \uparrow|)^T\nonumber\\
&\times H_{N Q}(|j,\psi_{n-1~q\mp(-1)^j}\rangle\otimes|\downarrow\rangle_N,|j,\psi_{N-1~q\pm(-1)^j}\rangle\otimes|\uparrow\rangle_N),\label{pp1}\\
O_{nj}^\pm=&(\langle j+1,\psi_{N-1~q\pm(-1)^{j}}|\otimes{}_N\langle
\downarrow|,\langle j+1,\psi_{N-1~q\mp(-1)^{j}}|\otimes{}_N\langle
\uparrow|)^T\nonumber\\
&\times H_{N
Q}(|j,\psi_{N-1~q\mp(-1)^j}\rangle\otimes|\downarrow\rangle_N,|j,\psi_{N-1~q\pm(-1)^j}\rangle\otimes|\uparrow\rangle_N)\label{pp2}
\end{eqnarray}
where
\begin{eqnarray} H_{NQ}=H_{N-1~
Q}+\Delta_N\sigma_{Nz}+g_N\sigma_{Nx}(a+a^\dagger).
\end{eqnarray}
Then we have
\begin{eqnarray}
D_{N~j}^\pm=\left(
\begin{array}{cc}
D_{N-1~j}^\mp-\Delta_N&0 \\
0&D_{N-1~j}^\pm+\Delta_N\\
\end{array}
\right),\\
\label{cond-1} O_{N~j}^\pm=O_{N~j}^\mp=O_{N~j}=\left(
\begin{array}{cc}
O_{N-1~j} &\sqrt{j+1}g_N I \\
\sqrt{j+1}g_N I&O_{N-1~j} \\
\end{array}
\right),
\end{eqnarray}
with the initial condition
\begin{eqnarray}
D_{1~j}^\pm=j\pm(-1)^j\Delta_1,\\
O_{1~j}^\pm=\sqrt{j+1}g_1.
\end{eqnarray}
As seen from Eq. \eqref{cond-1}, generally there is no closed
subspace if $O_{nj}$ is nontrival, which is exactly the case for
single qubit with $g\neq 0$. But for the multi-qubit case, $O_{Nj}$
can be equivalently trivial even for non-zero coupling constant
$g_i$, if its eigenvalues are $0$, which leads to the closed
subspaces. Suppose that a subspace is spanned by
$\{|J,\psi_{Nq\pm(-1)^J}\rangle,|J+1,\psi_{Nq\pm(-1)^{J+1}}\rangle,\ldots,|K,\psi_{Nq\pm}(-1)^K\rangle\}$.
If
\begin{eqnarray}
O_{N~J-1} ~c^\pm_{N,J}|J,\psi_{Nq\pm(-1)^J}\rangle=0,\label{r1}\\
O_{N~K}~c^\pm_{N,K}|K,\psi_{Nq\pm(-1)^K}\rangle=0,\label{r2}
\end{eqnarray}
where $c^\pm_{N,J}$ and $c^\pm_{N,K}$ are coefficients of
$|J,\psi_{Nq\pm(-1)^J}\rangle$ and $|K,\psi_{Nq\pm(-1)^K}\rangle$
respectively, then this subspace is closed. Each of $c^\pm_{N,J}$
and $c^\pm_{N,K}$ contains $2^{N-1}$ components since
$|J,\psi_{Nq\pm(-1)^J}\rangle$ and $|K,\psi_{Nq\pm(-1)^K}\rangle$
are $2^{N-1}$ dimensional vectors. So combined with the eigenvalue
equation of $H_{NQ}$ in this closed subspace, we obtain
\begin{eqnarray}\fl\footnotesize\label{eig}
\left(
  \begin{array}{cccccc}
   O_{N~J-1} &0 &0 & 0&\dots\\
   D_{NJ}^\pm-E^\pm &O_{N~J}&0 & 0&\dots\\
    O_{N~J} & D_{N~J+1}^\pm-E^\pm&O_{N~J+1}& 0 &\dots  \\
      \dots&\dots&\dots&\dots&\dots\\
      0&\dots&O_{N~K-2} &D_{NK-1}^\pm-E^\pm &O_{N~K-1} \\
   0&\dots&0&O_{N~K-1} &D_{NK}^\pm-E^\pm  \\
    0&\dots&0&0 & O_{NK} \\
  \end{array}
\right)\left(
  \begin{array}{c}
   c^\pm_{N,J} \\
   c^\pm_{N,J+1} \\
      \ldots\\
   c^\pm_{N,K-1}\\
   c^\pm_{N,K} \\
  \end{array}
\right)=0.
\end{eqnarray}
Clearly, there are more equations (rows) than variables (columns) in
this system of linear homogeneous equations, so only for some
special conditions, we may obtain a solution with finite photon
numbers. We can use elementary row transformations to reduce the
matrix into row echelon form, then if the number of the non-zero
rows is less than that of the columns, there will be non-trivial
solutions. At the same time, Eqs. \eqref{r1} and \eqref{r2} are
decoupled from other equations in Eq. \eqref{eig}, which is just the
existence condition of the closed subspace, and they differ only by
a constant. We can eliminant all the constants and define
$O_{N}=O_{NJ}/\sqrt{J+1}$, then Eqs. \eqref{r1} and \eqref{r2} can
be equivalent to the statement that the eigenvalues of $O_{n}$ are
zero, and both $c^\pm_{N,J}|J,\psi_{Nq\pm(-1)^J}\rangle$ and
$c^\pm_{N,K}|K,\psi_{Nq\pm(-1)^J}\rangle$ are its null vectors.

$O_{N}$ takes different forms for qubit number $N$, but we can find
a unified form for its eigenvalues by analyzing its determinant
\begin{eqnarray}
\left|O_{N}\right|&=\left|\left(
\begin{array}{cc}
O_{N-1} &g_N  \\
g_N &O_{N-1} \\
\end{array}\right)\right|=
\left|\left(
\begin{array}{cc}
O_{N-1} &g_N  \\
g_N-O_{N-1} &O_{N-1}-g_N \\
\end{array}\right)\right|\nonumber\\
&= \left|\left(
\begin{array}{cc}
O_{N-1}+g_N  &0 \\
g_N-O_{N-1} &O_{N-1}-g_N \\
\end{array}\right)\right|=
\left|O_{N-1}+g_N \right|\left|O_{N-1}-g_n \right|.
\end{eqnarray}
So if the eigenvalues of $O_{N-1}$ are
$e_{N-1,i}~(i=1,2,\ldots,2^{N-2})$, then the eigenvalues of $O_{N}$
would be $e_{N-1,i}+g_N$ and $e_{N-1,i}-g_N$ with the initial
condition $e_{1,1}=g_1$. Some eigenvalues and eigenvectors of
$O_{N}$ are shown in table \ref{t1}.
\begin{table}[htbp]
\centering\caption{\label{t1}Qubit number $N$, eigenvalues and
corresponding eigenvectors of $O_N$.}
\begin{tabular}{p{0.8cm}p{3.6cm}p{5.4cm}}
\hline \hline
$N$&eigenvalues&transpose of the eigenvectors\\
\hline
$2$&$g_1+g_2$&$(1,1)$\\
$2$&$g_1-g_2$&$(1,-1)$\\
$3$&$g_1 - g_2 - g_3$&$(1, -1, -1, 1)$\\
$3$&$g_1 + g_2 - g_3$&$(-1, -1, 1, 1)$\\
$3$&$g_1 - g_2 + g_3$&$(-1, 1, -1, 1)$\\
$3$&$g_1 + g_2 + g_3$&$(1, 1, 1, 1)$\\
$4$&$g_1 - g_2 - g_3 - g_4$&$(-1, 1, 1, -1, 1, -1, -1, 1)$\\
$4$&$g_1 + g_2 - g_3 - g_4$&$(1, 1, -1, -1, -1, -1, 1, 1)$\\
$4$&$g_1 - g_2 + g_3 - g_4$&$(1, -1, 1, -1, -1, 1, -1, 1)$\\
$4$&$g_1 + g_2 + g_3 - g_4$&$(-1, -1, -1, -1, 1, 1, 1, 1)$\\
$4$&$g_1 - g_2 - g_3 + g_4$&$(1, -1, -1, 1, 1, -1, -1, 1)$\\
$4$&$g_1 + g_2 - g_3 + g_4$&$(-1, -1, 1, 1, -1, -1, 1, 1)$\\
$4$&$g_1 - g_2 + g_3 + g_4$&$(-1, 1, -1, 1, -1, 1, -1, 1)$\\
$4$&$g_1 + g_2 + g_3 + g_4$&$(1,~ 1,~ 1,~ 1,~ 1,~ 1,~ 1,~ 1)$\\
\hline \hline
\end{tabular}
\end{table}

Setting the eigenvalues of $O_{n}$ to be $0$, which just depends on
the coupling strength, and
$c^\pm_{N,J}|J,\psi_{Nq\pm(-1)^J}\rangle$,
$c^\pm_{N,K}|K,\psi_{Nq\pm(-1)^J}\rangle$ to be its null vectors, we
can simplify \eqref{eig}. Now the relations between all the
components of each of $c^\pm_{N,J}$ and $c^\pm_{N,K}$ are fixed,
so there is only $1$ variable. Meanwhile, using
$O_{N,J-1}c^\pm_{N,J}=0$ and $O_{N,K}c^\pm_{N,K}=0$ to simplify Eq.
\eqref{eig} by elementary row transformation and then put them
aside, we obtain a necessary but not sufficient condition for a
solution
\begin{eqnarray}
\left|
  \begin{array}{cccccc}
   D_{N~J}^\pm-E^\pm &O_{N~J}&0 & 0&\dots\\
    0 & D_{N~J+1}^\pm-E^\pm&O_{N~J+1}& 0 &\dots  \\
      \dots&\dots&\dots&\dots&\dots\\
      0&\dots&O_{N~K-2} &D_{N~K-1}^\pm-E^\pm &0 \\
   0&\dots&0&O_{N~K-1} &D_{N~K}^\pm-E^\pm
  \end{array}
\right|=0,
\end{eqnarray}
which is generally dependent both on the qubit energy and coupling
strength, except for $K=J+1$. But if $J\neq0$, the corresponding
wavefunction will not depend on the coupling strength at all and it
turns into the ``dark state''. So in order to search for the
dark-like states, we just need to consider the case of $J=0$ and
$K=1$. The equations which determine the solution to $H_{NQ}$ reads
\begin{eqnarray}\fl\label{eig1}
\left(
  \begin{array}{cc}
   D_{N0}^\pm-E^\pm &0 \\
    O_{N~0} & D_{N~1}^\pm-E^\pm  \\
      0& O_{n~1} \\
  \end{array}
\right)\left(
  \begin{array}{c}
   c^\pm_{N,0} \\
   c^\pm_{N,1} \\
  \end{array}
\right)=0.
\end{eqnarray}
Solving this equation is the key point to obtaining the dark-like eigenstates for the multi-qubit and multi-photon Rabi models.

Let us start with the simplest case of $N=2$. For this case, we have
$(|\psi\rangle_{2q+})=(|\downarrow,\downarrow\rangle,|\uparrow,\uparrow\rangle)$,
$(|\psi\rangle_{2q-})=(|\uparrow,\downarrow\rangle,|\downarrow,\uparrow\rangle)$,
and
\begin{eqnarray}
O_{2}=\left(\begin{array}{cc}
                    g_1 & g_2 \\
                    g_2 & g_1
                  \end{array}\right),
                  \end{eqnarray}
whose eigensystem is shown in table 1. Choosing $g_1=g_2=g/2$ and
$c_{2,1,1}=-c_{2,1,2}$ to simplify \eqref{eig1}, we arrive at
\begin{eqnarray}\label{eig2a}\fl
\left(
  \begin{array}{cccc}
   \mp\Delta_1-\Delta_2-E^\pm &0 &0  \\
    0&\pm\Delta_1+\Delta_2-E^\pm &0 \\
      g/2& g/2&1\pm\Delta_1-\Delta_2-E^\pm \\
       g/2& g/2&-1\pm\Delta_1-\Delta_2+E^\pm \\
  \end{array}
\right)\left(
  \begin{array}{c}
   c^\pm_{2,0,1} \\
c^\pm_{2,0,2}\\
   c^\pm_{2,1,1} \\
  \end{array}
\right)=0.
\end{eqnarray}
After elementary row transformation, the coefficient matrix in Eq.
\eqref{eig2a} is simplified to the form
\begin{eqnarray}\label{eig3}
\left(
  \begin{array}{ccc}
    g/2& g/2&1\pm\Delta_1-\Delta_2-E^\pm \\
  \mp\Delta_1-\Delta_2-E^\pm &0 &0 \\
   0&\pm\Delta_1+\Delta_2-E^\pm &0 \\
       0&0&E^\pm-1 \\
  \end{array}
\right)
\end{eqnarray}
There are three columns, so only two non-zero rows can exist in its
row echelon form, from which we obtain
\begin{eqnarray}\label{eig4}
\Delta_1+\Delta_2=E^+=1,
\end{eqnarray}
with eigenstate
\begin{eqnarray}\label{dk1}
|\psi\rangle_{e}=\frac{1}{{\cal
N}}\left(\frac{2(\Delta_1-\Delta_2)}{g}|0,\uparrow,\uparrow\rangle-
 |1,\uparrow,\downarrow\rangle
+|1,\downarrow,\uparrow\rangle\right),
\end{eqnarray}
 for even parity and
\begin{eqnarray}\label{eig2}
\Delta_1-\Delta_2=E^-=1,~or~\Delta_2-\Delta_1=E^-=1
\end{eqnarray}
with eigenstates
\begin{eqnarray}
|\psi\rangle_{g1}=\frac{1}{{\cal
N}}\left(\frac{2(\Delta_1+\Delta_2)}{g}|0,\uparrow,\downarrow\rangle+
 |1,\downarrow,\downarrow\rangle
-|1,\uparrow,\uparrow\rangle\right),\label{dk2}\\
 |\psi\rangle_{g2}=\frac{1}{{\cal
N}}\left(\frac{2(\Delta_1+\Delta_2)}{g}|0,\downarrow,\uparrow\rangle+
 |1,\downarrow,\downarrow\rangle
-|1,\uparrow,\uparrow\rangle\right),\label{dk3}
\end{eqnarray}
respectively, for odd parity. Eigenstates \eqref{dk1}, \eqref{dk2},
\eqref{dk3} exist for any coupling strength $g_1=g_2=g/2$ with
constant eigenenergy $E^\pm=1$, corresponding to a horizontal line
in the spectra, which has been shown in Ref. \cite{pj2}. These
properties are just like those for the ``dark state'' formed by the
qubit singlet. However, for these eigenstates, the qubit and photon
are not decoupled, and the photon number is bounded from above at
$1$.

Then we consider the case of $3$ qubit, where
$(|\psi\rangle_{3q+})=(|\downarrow,\downarrow,\downarrow\rangle,|\uparrow,\uparrow,\downarrow\rangle,|\uparrow,\downarrow,\uparrow\rangle,|\downarrow,\uparrow,\uparrow\rangle)$,
$(|\psi\rangle_{3q-})=(|\uparrow,\downarrow,\downarrow\rangle,|\downarrow,\uparrow,\downarrow\rangle,|\downarrow,\downarrow,\uparrow\rangle,|\uparrow,\uparrow,\uparrow\rangle)$,
and
\begin{eqnarray}
O_{3}=\left(\begin{array}{cccc}
                    g_1 & g_2&g_3&0 \\
                    g_2 & g_1&0&g_3\\
                    g_3 & 0&g_1&g_2 \\
                    0 & g_3&g_2&g_1\\
                  \end{array}\right),
                  \end{eqnarray}
whose eigensystem is shown in table 1. For $g_1=g_2+g_3$,
$g_2=g_1+g_3$, or $g_3=g_1+g_2$, the eigenvalues are zero, and
corresponding eigenvectors are nullvectors.
$(D_{3~0}^\pm-E^\pm)c^\pm_{3,0}=0$ in Eq. \eqref{eig1} is decoupled from other
parts. The coefficient matrix $D_{3~0}^\pm-E^\pm$ takes the
diagonal form
\begin{eqnarray}\label{eig7}\fl\scriptsize
\left(
  \begin{array}{ccccc}
   \pm\Delta_1-\Delta_2-\Delta_3-E^\pm &0 &0&0  \\
    0&\mp\Delta_1+\Delta_2-\Delta_3-E^\pm &0&0 \\
      0& 0&\mp\Delta_1-\Delta_2+\Delta_3-E^\pm&0 \\
       0& 0&0&\pm\Delta_1+\Delta_1+\Delta_3-E^\pm\\
       \end{array}\right).
       \end{eqnarray}
 Choosing $g_1=g_2+g_3$ and
$c_{3,1,1}=-c_{3,1,2}=-c_{3,1,3}=c_{3,1,4}$ to simplify the other part
\begin{eqnarray}\fl\label{eig8}
\left(
  \begin{array}{cc}
    O_{3~0} & D_{3~1}^\pm-E^\pm  \\
      0& O_{3~1} \\
  \end{array}
\right)\left(
  \begin{array}{c}
   c^\pm_{3,0} \\
   c^\pm_{3,1} \\
  \end{array}
\right)=0,
\end{eqnarray}
we arrive at
\begin{eqnarray}\label{eig9}\fl\small
\left(
  \begin{array}{ccccc}
  g_2+g_3 & g_2&g_3&0&1\mp\Delta_1-\Delta_2-\Delta_3-E^\pm \\
   g_2 & g_2+g_3&0&g_3&-(1\pm\Delta_1+\Delta_2-\Delta_3-E^\pm)\\
   g_3 & 0&g_2+g_3&0&-(1\pm\Delta_1-\Delta_2+\Delta_3-E^\pm) \\
   0 & g_3&0&g_2+g_3&1\mp\Delta_1+\Delta_1+\Delta_3-E^\pm\\
  \end{array}
\right)\left(
  \begin{array}{c}
   c^\pm_{3,0,1} \\
c^\pm_{3,0,2}\\
   c^\pm_{3,0,3} \\
   c^\pm_{3,0,4}\\
   c^\pm_{3,1,1} \\
  \end{array}
\right)=0.
\end{eqnarray}
After elementary row transformation, the coefficient matrix in Eq.
\eqref{eig9} becomes
\begin{eqnarray}\label{eig100}\small\fl \left(
  \begin{array}{ccccc}
1 & 0  & 0 &  -1 & \frac{1-\Delta_3-E^\pm}{ g_3}+\frac{1-\Delta_2-E^\pm}{g_2}\\
0 & 1 &0 &1
&\frac{-1+\Delta_3+E^\pm}{g_3}+\frac{-1\mp\Delta_1+E^\pm}{g_2+g_3}\\
 0& 0& 1& 1&
\frac{-1+\Delta_2+E^\pm}{g_2}+\frac{-1\mp\Delta_1+E^\pm}{g_2+g_3}\\
 0 &  0&   0&   0&   4-4 E^\pm\\
  \end{array}
\right)
\end{eqnarray}
There are totally $5$ variables, so the total nonzero rows in Eqs.
\eqref{eig7} and \eqref{eig100} should be less than $5$. By choosing
$E^\pm=1$, the nonzero rows in Eq. \eqref{eig100} reduce to $3$,
which means only one nonzero row can exist in Eq. \eqref{eig7} to
obtain a nontrivial solution. Luckily, there is one such case for
odd parity with the following parameters
\begin{eqnarray}\label{1}
\Delta_1=\Delta_2=\Delta_3=\Delta=E^-=1.
\end{eqnarray}
Substituting Eq. \eqref{1} into Eqs. \eqref{eig7} and
\eqref{eig100}, we obtain a dark-like state
\begin{eqnarray}\label{2}
|\psi\rangle=&\frac{g_3}{g_2(g_2+g_3)}|0,\uparrow,\uparrow,\downarrow\rangle+\frac{g_2}{g_3(g_2+g_3)}|0,\uparrow,\downarrow,\uparrow\rangle
-\frac{g_2+g_3}{g_2g_3}|0,\downarrow,\uparrow,\uparrow\rangle\nonumber\\
&+|1,\uparrow,\downarrow,\downarrow\rangle-|1,\downarrow,\uparrow,\downarrow\rangle
-|1,\downarrow,\downarrow,\uparrow\rangle+|1,\uparrow,\uparrow,\uparrow\rangle
\end{eqnarray}

If we choose $g_2=g_1+g_3$ and
$-c_{3,1,1}=c_{3,1,2}=-c_{3,1,3}=c_{3,1,4}$ to satisfy the condition
$O_{3~1}c^\pm_{3,1}=0$, the corresponding solution can be obtained
just by interchanging the states of the first and second qubits,
including the coupling strength in Eq. \eqref{2}. For $g_3=g_1+g_2$,
we can get a solution by interchanging the states of the first and
third qubit in Eq. \eqref{2}. Choosing $g_1=g_2+g_3$ and
$\Delta_1=\Delta_2=\Delta_3=\Delta=1$, the dark-like state \eqref{2}
corresponds to the horizontal line $E^-=1$ in Figure 1.

\begin{figure}[htbp]
\center
\includegraphics[width=0.8\textwidth]{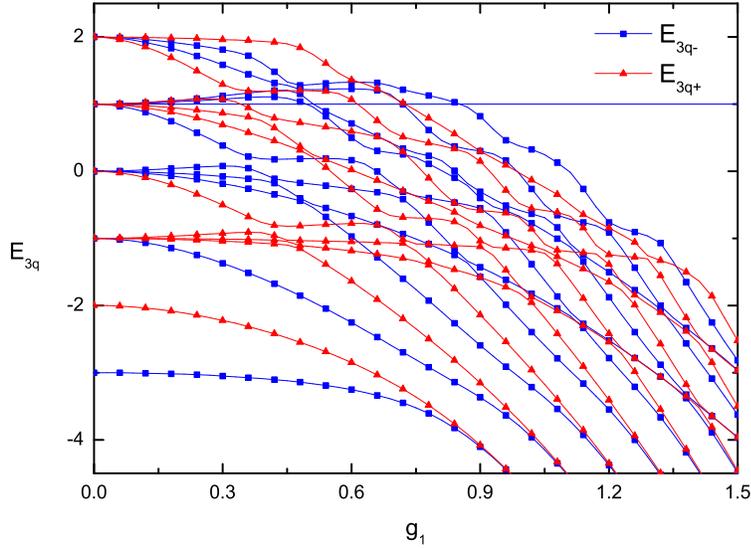}
\renewcommand\figurename{\textbf{Figure}}
\caption[2]{The numerical spectrum of three-qubit quantum Rabi model
with
$\Delta_1=\Delta_2=\Delta_3=1$,~$\omega=1$,~$g_3=0.5g_2$,~$0~\leq~
g_1=g_2+g_3~\leq~1.5$. $E_{+}$ and $E_{-}$ are solutions with even
and odd parity respectively.\label{figure1}}
\end{figure}

Now we turn to the case of $4$ qubit, with
$(|\psi\rangle_{4q+})=(|\psi\rangle_{3q-}\otimes|\downarrow\rangle_4,|\psi\rangle_{3q+}\otimes|\uparrow\rangle_4)$,
$(|\psi\rangle_{4q-})=(|\psi\rangle_{3q+}\otimes|\downarrow\rangle_4,|\psi\rangle_{3q-}\otimes|\uparrow\rangle_4)$,
and
\begin{eqnarray}
O_{4}=\left(\begin{array}{cccccccc}
                    g_1 & g_2&g_3&0 &g_4&0&0&0\\
                    g_2 & g_1&0&g_3&0&g_4&0&0\\
                    g_3 & 0&g_1&g_2&0&0&g_4&0 \\
                    0 & g_3&g_2&g_1&0&0&0&g_4\\
                    g_4 &0&0&0& g_1 & g_2&g_3&0\\
                    0&g_4 &0&0& g_2 & g_1&0&g_3\\
                    0 &0&g_4&0& g_3 & 0&g_1&g_2\\
                    0 &0&0&g_4& 0 & g_3&g_2&g_1\\
                  \end{array}\right),
                  \end{eqnarray}
As seen from the system of linear homogeneous equations
\eqref{eig1}, there are $24$ rows and just $16$ columns in its
coefficient matrix. A solution exists if the nonzero rows is less
than the columns in its row echelon form. First we consider the
$8\times 8$ diagonal matrix in \eqref{eig1}
\begin{eqnarray}\label{d4}
D_{4~0}^\pm-E^\pm=\left(
\begin{array}{cc}
D_{3~0}^\mp-\Delta_4-E^\pm&0 \\
0&D_{3~0}^\pm+\Delta_4-E^\pm\\
\end{array}
\right),
\end{eqnarray}
and then take into account the other part
\begin{eqnarray}
\label{eig16} \left(
  \begin{array}{cc}
    O_{4~0} & D_{4~1}^\pm-E^\pm  \\
      0& O_{4~1} \\
  \end{array}
\right)\left(
  \begin{array}{c}
   c^\pm_{4,0} \\
   c^\pm_{4,1} \\
  \end{array}
\right)=0.
\end{eqnarray}
We can simplify the condition $O_{4~1}  c^\pm_{4,1}=0$ by setting
one of the eigenvalues of $O_{4}$ to be $0$ and $(c^\pm_{4,1})$ to
be its null vector (shown in table 1). This will eliminate $8$ rows
and $7$ columns in the coefficient matrix in Eq. \eqref{eig16}. If
all the coupling strengths $g_i$ $(i=1,2,3,4)$ are nonzero, then
there are at least $7$ nonzero rows in the row echelon form of this
coefficient matrix, because the number of the zero rows in the
echelon form of $O_{4~0}$ is just the same as its null vectors.
Together with the diagonal matrix $D_{4~0}^\pm-E^\pm$, there are at
least $15$ rows but only $9$ columns totally, so there should be at
least $7$ zero rows in $D_{4~0}^\pm-E^\pm$, which is impossible by
analyzing Eq. \eqref{d4}.

 It seems that there are no dark-like solutions for the 4-qubit Rabi
 model up to now. However, there are other possibilities by setting more than $1$ eigenvalues in table 1 to be $0$ simultaneously,
  which will eliminant more rows and less columns because there are more null vectors for
 $O_4$. By analyzing table 1, we can choose $g_1=g_2$ and $g_3=g_4$
 to set the eigenvalues $g_1 - g_2 + g_3 - g_4$ and $g_1 - g_2 - g_3 +
 g_4$ to be $0$ simultaneously, then $(c^\pm_{4,1}|\psi_{4q\mp}\rangle)$
 can be the linear superposition of the corresponding two null vectors (shown in Tab. 1), and there will
 be two variables. After elementary row transformation, the coefficient matrix in Eq. \eqref{eig16} reduces to row echelon
 form
 \begin{eqnarray}\fl\scriptsize\displaystyle\label{4q}
 \left(\begin{array}{cccccccccc}
 1&  0 &  0 &  0&   0&   0&   0&   -1&  \frac{1\pm\Delta_1-\Delta_2-E^\pm}{g_1}+\frac{1-\Delta_3+\Delta_4-E^\pm}{g_3}&
 \frac{1\pm\Delta_1-\Delta_2-E^\pm}{g_1}+\frac{\Delta_3-\Delta_4+E^\pm-1}{g_3}\\
0&   1&   0 &  0&   0&   0&   0&   1&
\frac{-1+\Delta_3-\Delta_4+E^\pm}{g_3}+\frac{E^\pm-1}{g_1-g_3}& \frac{1-\Delta_3+\Delta_4-E^\pm}{g_3}+\frac{E^\pm-1}{g_1+g_3}\\
0& 0& 1& 0&   0&   -1&  0&   0&
\frac{1-\Delta_3-\Delta_4-E^\pm}{g_3}+\frac{1\pm\Delta_1+\Delta_2-E^\pm}{g_1}&
\frac{1-\Delta_3-\Delta_4-E^\pm}{g_3}-\frac{1\pm\Delta_1+\Delta_2-E^\pm} {g_1}\\
0&   0 &  0&   1&   0&   1& 0& 0&
\frac{-1+\Delta_3+\Delta_4+E^\pm}{g_3}+\frac{E^\pm-1}{g_3-g_1}&
\frac{-1+\Delta_3+\Delta_4+E^\pm}{g_3}+\frac{E^\pm-1}{g_1+g_3}\\
0& 0& 0& 0&   1&   1&   0&   0& \frac{-1\mp\Delta_1-\Delta_2+E^\pm}
{g_1}+\frac{E^\pm-1}{g_3-g_1}& \frac{1\pm\Delta_1+\Delta_2-E^\pm}{g_1}+\frac{E^\pm-1}{g_1+g_3}\\
0&   0&   0& 0&   0& 0& 1&   1&
\frac{-1\mp\Delta_1+\Delta_2+E^\pm}{g_1}+\frac{E^\pm-1}{g_1-g_3}&
\frac{-1\mp\Delta_1+\Delta_2+E^\pm}{g_1}+\frac{E^\pm-1}{g_1+g_3}\\
 0&   0& 0&   0&   0&   0&   0&   0&   4E^\pm-4&    4-4E^\pm\\
 0&   0&   0&   0&   0& 0& 0& 0&   4-4E^\pm&   4-4E^\pm\\
 \end{array}
 \right).\end{eqnarray}
If $E^\pm=1$, Eq. \eqref{4q} reduces to
\begin{eqnarray}\fl\small\displaystyle\label{4q17}
 \left(\begin{array}{cccccccccc}
 1&  0 &  0 &  0&   0&   0&   0&   -1&  \frac{\pm\Delta_1-\Delta_2}{g_1}+\frac{-\Delta_3+\Delta_4}{g_3}&
 \frac{\pm\Delta_1-\Delta_2}{g_1}+\frac{\Delta_3-\Delta_4}{g_3}\\
0&   1&   0 &  0&   0&   0&   0&   1&
\frac{\Delta_3-\Delta_4}{g_3}& \frac{-\Delta_3+\Delta_4}{g_3}\\
0& 0& 1& 0&   0&   -1&  0&   0&
-\frac{\Delta_3+\Delta_4}{g_3}+\frac{\pm\Delta_1+\Delta_2}{g_1}&
\frac{-\Delta_3-\Delta_4}{g_3}-\frac{\pm\Delta_1+\Delta_2} {g_1}\\
0&   0 &  0&   1&   0&   1& 0& 0& \frac{\Delta_3+\Delta_4}{g_3}&
\frac{\Delta_3+\Delta_4}{g_3}\\
0& 0& 0& 0&   1&   1&   0&   0& -\frac{\pm\Delta_1+\Delta_2}
{g_1}& \frac{\pm\Delta_1+\Delta_2}{g_1}\\
0&   0&   0& 0&   0& 0& 1&   1& \frac{\mp\Delta_1+\Delta_2}{g_1}&
\frac{\mp\Delta_1+\Delta_2}{g_1}\\
 0&   0& 0&   0&   0&   0&   0&   0&   0&    0\\
 0&   0&   0&   0&   0& 0& 0& 0&   0&  0\\
 \end{array}
 \right),\end{eqnarray}
then together with $D_{4~0}^\pm-E^\pm$, there are $14$ rows but just
$10$ columns totally, so there should be $5$ zero rows in
 $D_{4~0}^\pm-E^\pm$, which seems impossible by only analyzing Eq. \eqref{d4},
 but this is indeed not the case. For even parity, if
 $\Delta_1-\Delta_2=\pm1=\pm E$ and  $\Delta_3=\Delta_4$, there are dark-like state solutions by
 analyzing Eqs. \eqref{4q17} and \eqref{d4},
\begin{eqnarray}
|\psi\rangle_{g1}=\frac{1}{{\cal
N}}\left(\frac{2(\Delta_1+\Delta_2)}{g}|0,\uparrow,\downarrow\rangle+
 |1,\downarrow,\downarrow\rangle
-|1,\uparrow,\uparrow\rangle\right)\otimes(|\uparrow\downarrow-|\downarrow\uparrow),\label{dk202}\\
 |\psi\rangle_{g2}=\frac{1}{{\cal
N}}\left(\frac{2(\Delta_1+\Delta_2)}{g}|0,\downarrow,\uparrow\rangle+
 |1,\downarrow,\downarrow\rangle
-|1,\uparrow,\uparrow\rangle\right)\otimes(|\uparrow\downarrow-|\downarrow\uparrow),\label{dk23}
\end{eqnarray}
where the first two qubits form a two-qubit dark-like state
\eqref{dk2} and \eqref{dk3} respectively, and another two qubits form a spin singlet
dark state. For odd parity, if $\Delta_1+\Delta_2=1=E^-$, there are
similar dark-like states formed by
$|\psi\rangle_{e}\otimes(|\uparrow\downarrow-|\downarrow\uparrow)$,
where $|\psi\rangle_{e}$ is given by Eq. \eqref{dk1}.

 If $E^\pm\neq1$, then Eq. \eqref{4q} reduces to
\begin{eqnarray}\fl\small\displaystyle\label{4q1}
 \left(\begin{array}{cccccccccc}
 1&  0 &  0 &  0&   0&   0&   0&   -1&  0&
0\\
0&   1&   0 &  0&   0&   0&   0&   1&
0& 0\\
0& 0& 1& 0&   0&   -1&  0&   0& 0&
0\\
0&   0 &  0&   1&   0&   1& 0& 0& 0&
0\\
0& 0& 0& 0&   1&   1&   0&   0&0& 0\\
0&   0&   0& 0&   0& 0& 1&   1& 0&
0\\
 0&   0& 0&   0&   0&   0&   0&   0&   1&    0\\
 0&   0&   0&   0&   0& 0& 0& 0&   0&  1\\
 \end{array}
 \right).\end{eqnarray}
For even parity, if $\Delta_1=\Delta_2$ and $\Delta_3=\Delta_4$,
there is a ``dark state'' solution
\begin{eqnarray}
|\psi\rangle_{d}=|0,\uparrow,\downarrow\rangle-
 |0,\downarrow,\uparrow\rangle
)\otimes(|\uparrow\downarrow-|\downarrow\uparrow),\label{ddk22}
\end{eqnarray}
which is just the product of the two-qubit singlet.

Finally, we come to the case $g_1=g_2=g_3=g_4=g$. Now three
eigenvalues $g_1 - g_2 + g_3 - g_4$, $g_1 - g_2 - g_3 +
 g_4$ and $g_1 +g_2 - g_3 - g_4$ are set to be $0$, and there are three
 null vectors shown in table 1, which can be simplified to $(1,0,0,-1,-1,0,0,1)^T$, $(0,1,0,-1,-1,0,1,0)^T$, $(0,0,1,-1,-1,1,0,0)^T$.
 Supposing that $(c^\pm_{4,1}|\psi_{4q\mp}\rangle)$ is the linear superposition of these null vectors, after elementary row transformation, the coefficient matrix in Eq. \eqref{eig16} reduces to row echelon  form
\begin{eqnarray}\fl\scriptsize\displaystyle\label{4qA}
 \left(\begin{array}{ccccccccccc}
 1&  0 &  0 &  0&   0&   0&   0&   -1&  \frac{3\pm\Delta_1-\Delta_4-3E^\pm}{g}&\frac{2+\Delta_2-\Delta_4-2E^\pm}{g}&
 \frac{5+\Delta_3-\Delta_4-5E^\pm}{g}\\
0&   1&   0 &  0&   0&   0&   -1&   0&
\frac{2-\Delta_2-\Delta_3-2E^\pm}{g}& \frac{1\mp\Delta_1-\Delta_3-E^\pm}{g}&\frac{3-\Delta_3+\Delta_4-3E^\pm}{g}\\
0& 0& 1& 0&   0&   -1&  0&   0&
\frac{2-\Delta_2-\Delta_3-2E^\pm}{g}&\frac{2-\Delta_2+\Delta_4-2E^\pm}{g}&
\frac{2\mp\Delta_1-\Delta_2-2E^\pm}{g}\\
0&   0 &  0&   1&   0&   1& 1& 1&
\frac{-3\mp\Delta_1+\Delta_2+\Delta_3+\Delta_4+3E^\pm}{g}&\frac{2E^\pm-2}{g}&
\frac{2E^\pm-2}{g}\\
0& 0& 0& 0&   1&   1&   1&   1&
\frac{-2+\Delta_2+\Delta_3+2E^\pm}{g}&\frac{-1\pm\Delta_1+\Delta_3+E^\pm}{g}&
\frac{-2\pm\Delta_1+\Delta_2+2E^\pm}{g}\\
0&   0&   0& 0&   0& 0& 0&   0&
\frac{2-2E^\pm}{g}&\frac{4-4E^\pm}{g}&
\frac{2-2E^\pm}{g}\\
 0&   0& 0&   0&   0&   0&   0&   0&  0&\frac{2E^\pm-2}{g}&\frac{2-2E^\pm}{g}\\
 0&   0&   0&   0&   0& 0& 0& 0& 0& 0&   \frac{8E^\pm-8}{g}\\
 \end{array}
 \right).\end{eqnarray}
If $E^\pm=1$, Eq. \eqref{4qA} reduces to
\begin{eqnarray}\fl\scriptsize\displaystyle\label{4qAwa}
 \left(\begin{array}{ccccccccccc}
 1&  0 &  0 &  0&   0&   0&   0&   -1&  \frac{\pm\Delta_1-\Delta_4}{g}&\frac{\Delta_2-\Delta_4}{g}&
 \frac{\Delta_3-\Delta_4}{g}\\
0&   1&   0 &  0&   0&   0&   -1&   0&
\frac{-\Delta_2-\Delta_3}{g}& \frac{\mp\Delta_1-\Delta_3}{g}&\frac{-\Delta_3+\Delta_4}{g}\\
0& 0& 1& 0&   0&   -1&  0&   0&
\frac{-\Delta_2-\Delta_3}{g}&\frac{-\Delta_2+\Delta_4}{g}&
\frac{\mp\Delta_1-\Delta_2}{g}\\
0&   0 &  0&   1&   0&   1& 1& 1&
\frac{\mp\Delta_1+\Delta_2+\Delta_3+\Delta_4}{g}&0&
0\\
0& 0& 0& 0&   1&   1&   1&   1&
\frac{\Delta_2+\Delta_3}{g}&\frac{\pm\Delta_1+\Delta_3}{g}&
\frac{\pm\Delta_1+\Delta_2}{g}\\
0&   0&   0& 0&   0& 0& 0&   0& 0&0&
0\\
 0&   0& 0&   0&   0&   0&   0&   0&  0&0&0\\
 0&   0&   0&   0&   0& 0& 0& 0& 0& 0&   0\\
 \end{array}
 \right).
 \end{eqnarray}
then together with $D_{4~0}^\pm-E^\pm$, there are $13$ rows but just
$11$ columns totally, so there should be $3$ zero rows in
 $D_{4~0}^\pm-E^\pm$, which is possible by chosing
\begin{eqnarray}
\Delta_2=\Delta_3=\Delta_4=\pm\Delta_1-1~~~or\label{3e1}\\
\Delta_2=\Delta_3=\Delta_4=\pm\Delta_1+1.\label{3e2}
\end{eqnarray}
We can interchange $\Delta_2$ with $\Delta_1$, $\Delta_3$ and
$\Delta_4$, so there are totally $8$ choices, but we only consider
\eqref{3e1} and \eqref{3e2}, because they are equivalent.

Substituting Eq. \eqref{3e1} into Eq. \eqref{4qAwa}, for even
parity, we obtain one dark-like eigenstate
 \begin{eqnarray}\fl\label{3dark}
 &|\psi\rangle_{g1}=a\left(\frac{(\Delta_1+\Delta_2)}{g}|0,\uparrow,\downarrow\rangle+
 |1,\downarrow,\downarrow\rangle
-|1,\uparrow,\uparrow\rangle\right)_{1,2}\otimes(|\uparrow\downarrow-|\downarrow\uparrow)_{3,4}\nonumber\\&+b
\left(\frac{(\Delta_1+\Delta_2)}{g}|0,\uparrow,\downarrow\rangle+
 |1,\downarrow,\downarrow\rangle
-|1,\uparrow,\uparrow\rangle\right)_{1,3}\otimes(|\uparrow\downarrow-|\downarrow\uparrow)_{2,4}\nonumber\\&+c
\left(\frac{(\Delta_1+\Delta_2)}{g}|0,\uparrow,\downarrow\rangle+
 |1,\downarrow,\downarrow\rangle
-|1,\uparrow,\uparrow\rangle\right)_{1,4}\otimes(|\uparrow\downarrow-|\downarrow\uparrow)_{2,3}.
\end{eqnarray}
This can be easily understood due to the fact that for
$\Delta_2=\Delta_3=\Delta_4=\Delta_1-1$, there are three independent
solutions, each formed by the product of a two-qubit dark-like state
and a two-qubit singlet. For
$\Delta_2=\Delta_3=\Delta_4=\Delta_1+1$, the solution takes the same
form as \eqref{3dark} with the dark-like state substituted by
\eqref{dk3}. For even parity, we choose
$\Delta_2=\Delta_3=\Delta_4=-\Delta_1+1$, and the dark-like state
takes the same form as \eqref{3dark} with the dark-like state
substituted by \eqref{dk1}. Choosing $\Delta_1=\Delta_2+1$,
$\Delta_3=\Delta_4$, $g_1=g_2$, $g_3=g_4$, a dark-like state
\eqref{dk202} corresponding to the horizontal line $E^+=1$   is
shown in Figure 2.

\begin{figure}[htbp]
\center
\includegraphics[width=0.8\textwidth]{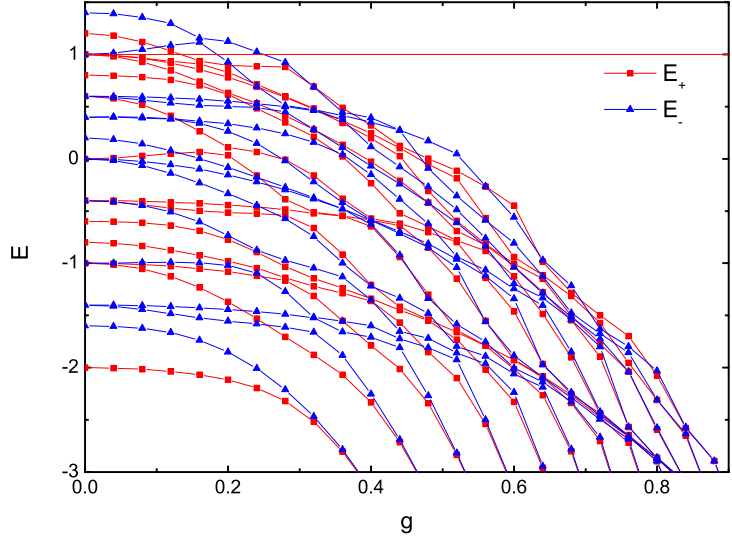}
\renewcommand\figurename{\textbf{Figure}}
\caption[2]{The numerical spectrum of the four-qubit quantum Rabi
model with
$\Delta_1=\Delta_2+1=1.2$,~$\Delta_3=\Delta_4=0.3$,~$\omega=1$,~$g_1=g_2=g_3=g_4$,~$0~\leq~
g_1=g_2=g_3=g_4=g~\leq~1$. $E_{+}$ and $E_{-}$ are solutions with
even and odd parity respectively.\label{figure1}}
\end{figure}

We can follow the similar procedure to find dark-like states for
$5,6,7,\ldots$ qubits cases. The key point is just to solve
\eqref{eig1} with different format. It should be pointed out that we
haven't found a universal existence condition and all the dark-like
states for arbitrary qubit number N, because it still needs detailed
analysis for more qubits. But we find one kind of dark-like states
commonly exist for arbitrary qubit number $N>1$
\begin{eqnarray}
|\psi\rangle_{Nq dark-like}=|\psi\rangle_{2q dark-like}\times(|\psi\rangle_{singlet})^{(N-2)/2}~~~~~~~N=2,4,6,8,\ldots\label{Nq1}\\
|\psi\rangle_{Nq dark-like}=|\psi\rangle_{3q
dark-like}\times(|\psi\rangle_{singlet})^{(N-3)/2}~~~~~~~N=3,5,7,9,\ldots
\end{eqnarray}
$N=4$ is an example of Eq. \eqref{Nq1}, and all its eigenstaes have
the form of Eq. \eqref{Nq1}.

\section{Dark-like states for the multi-qubit and multi-photon Rabi
model}\label{s3}

The N-qubit and M-photon Rabi model reads
\begin{equation}\label{gtq}
H_{PQ}=\omega
a^{\dagger}a+\sum_{i=1}^{N}g_{i}\sigma_{ix}(a^M+a^{\dagger^M})
+\sum_{i=1}^{N}\Delta_i\sigma_{iz},
\end{equation}
where $M$ is a positive integer. This model is of considerable
interest because of its relevance to the study of the coupling
between multi-qubit and photon field with the qubit making M-photon
transitions. Besides, it is known that under rotating wave
approximation, the dynamics of the M-photon J-C model \cite{xu} is
qualitatively different from that of the usual single-photon case
\cite{tp1}. As discussed in Ref. \cite{tp1,tp2,tp3}, for single
qubit case, this model is solvable only if $M\leq2$ and the coupling
parameter is below a certain critical value. But in the following
discussion, we will show that the case for more qubits is different:
Dark-like eigenstates for $H_{PQ}$ \eqref{gtq} with $N>1$ still
exist, regardless of these constraints, although in usual cases this
model is indeed not well-defined.

However, we first try to find out this critical value for $M=2$. We
assume that $\Delta_k(k=1,2,\ldots,N)=0$, which does not affect the
result \cite{tp2}. In the basis formed by the eigenstates of
$\prod\sigma_{jx}(j=1,2,\ldots,N)$, the Hamiltonian \eqref{gtq} with
$M=2$ is turned into the form \cite{tp2}
\begin{equation}\label{gtq1}
h_{PQ}=a^{\dagger}a+\lambda(a^M+a^{\dagger^M}),
\end{equation}
where $\lambda=\pm g_1\pm g_2\ldots\pm g_N$. Defining operators
\begin{eqnarray}
x=\frac{1}{\sqrt{2}}(a+a^\dagger),~~
p=i\sqrt{\frac{1}{2}}(a^\dagger-a),
\end{eqnarray}
then $h_{PQ}$ can be rewritten as
\begin{eqnarray}
h_{PQ}=p^2+\frac{1+2\lambda}{1-2\lambda}x^2-\frac{1}{2}.
\end{eqnarray}
Clearly, if $\frac{1+2\lambda}{1-2\lambda}=\omega^2>0$, then
$h_{PQ}$ corresponds to a quantum harmonic oscillator and can be
diagonalized. However, if
$\frac{1+2\lambda}{1-2\lambda}=-\omega^2<0$, $h_{PQ}$ represents an
inverted quantum harmonic oscillator, which cannot be diagonalized
using the basis states $|n\rangle$ of the number operator because
its eigenstates are not normalizable. Thus the condition for the
Hamiltonian \eqref{gtq1} being diagonalizable is
$\lambda<\frac{1}{2}$ \cite{tp2}, and correspondingly we have
$\max\{\pm g_1\pm g_2\ldots\pm g_N\}<\frac{1}{2}$, that is
\begin{equation}
\sum_{k=1}^{N}g_{k}<\frac{1}{2}.
\end{equation}

Then we search for the dark-like eigenstaes for $H_{PQ}$
\eqref{gtq}. There are $2M$ invariant subspaces
\begin{eqnarray}
\{|0,\psi_{Nq+}\rangle,|M,\psi_{Nq-}\rangle,|2M,\psi_{Nq+}\rangle,\ldots\}\nonumber\\
\{|0,\psi_{Nq-}\rangle,|M,\psi_{Nq+}\rangle,|2M,\psi_{Nq-}\rangle,\ldots\}\nonumber
\\
\{|1,\psi_{Nq+}\rangle,|M+1,\psi_{Nq-}\rangle,|2M+1,\psi_{Nq+}\rangle,\ldots\}\nonumber
\\
\{|1,\psi_{Nq-}\rangle,|M+1,\psi_{Nq+}\rangle,|2M+1,\psi_{Nq-}\rangle,\ldots\}\\
\ldots\nonumber\\
\{|M-1,\psi_{Nq+}\rangle,|2M-1,\psi_{Nq-}\rangle,|3M-1,\psi_{Nq+}\rangle,\ldots\}\nonumber\\
\{|M-1,\psi_{Nq-}\rangle,|2M-1,\psi_{Nq+}\rangle,|3M-1,\psi_{Nq-}\rangle,\ldots\},\nonumber
\end{eqnarray}
each of which can be labeled by $\{i,\pm\}$, where the initial
photon number takes the values $i=0,1,2,\ldots,M-1$, and $\pm$ is
the eigenvalue of $\prod_{k=1}^{N}\sigma_{kz}$ for the initial qubit
state. $H_{PQ}$ in each subspace has the same form as $H_{NQ}$
\eqref{eig1} except for some constants
\begin{eqnarray}\label{off}\fl\scriptsize\displaystyle
H_{PQi}^\pm=\left(
  \begin{array}{cccccc}
   i+D_{n0}^\pm &\sqrt{\frac{(i+M)!}{i!}}O_{n0}^\pm &0 & 0&0&\dots\\
   \sqrt{\frac{(i+M)!}{i!}} O_{n0}^\pm &i+M-1+ D_{n1}^\pm &\sqrt{\frac{(i+2M)!}{2(i+M)!}}O_{n1}^\pm & 0 &0&\dots  \\
   0&\sqrt{\frac{(i+2M)!}{2(i+M)!}}O_{n1}^\pm &i+2M-2+D_{n2}^\pm & \sqrt{\frac{(i+3M)!}{3(i+2M)!}}O_{n2}^\pm &0&\dots \\
     \dots&\dots&\dots&\dots&\dots&\dots\\
  \end{array}
\right),
\end{eqnarray}
where $D_{nj}^\pm$ and $O_{nj}^\pm$ ($j=0,1,2,3,\ldots$) are just
the same as defined in the N-qubit Rabi model in \eqref{d} and
\eqref{o}, respectively.

Now, to find out the dark-like solution, we follow the steps for the
N-qubit case to get
\begin{eqnarray}\fl\label{eig10}
\left(
  \begin{array}{cc}
   i+D_{n0}^\pm-E_{i}^\pm &0 \\
   \sqrt{\frac{(i+M)!}{i!}} O^\pm_{n~0} &i+M-1+ D_{n~1}^\pm-E_{i}^\pm  \\
      0& \sqrt{\frac{(i+M)!}{i!}} O^\pm_{n~1} \\
  \end{array}
\right)\left(
  \begin{array}{c}
   c^\pm_{n,0} \\
   c^\pm_{n,1} \\
  \end{array}
\right)=0.
\end{eqnarray}
If we define $E^\pm=E_{i}^\pm-(i+M-1)$ and
$g_{k}=\sqrt{\frac{i!}{(i+M)!}}g_{i,k}$ $(k=1,2,\ldots,N)$, so that
$\sqrt{\frac{(i+M)!}{i!}} O^\pm_{n~0,1}\rightarrow O^\pm_{n~0,1}$,
then we obtain
\begin{eqnarray}\fl\label{eig11}
\left(
  \begin{array}{cc}
   D_{n0}^\pm-(E^\pm+M-1) &0 \\
   O^\pm_{n~0} &D_{n~1}^\pm-E^\pm  \\
      0& O^\pm_{n~1} \\
  \end{array}
\right)\left(
  \begin{array}{c}
   c^\pm_{n,0} \\
   c^\pm_{n,1} \\
  \end{array}
\right)=0.
\end{eqnarray}
Eq. \eqref{eig11} has exactly the same form as Eq. \eqref{eig1},
except for $[D_{n0}^\pm-(E^\pm+M-1)]c^\pm_{n,0} =0$, which will just
determine the relation between $\Delta_k(k=1,2,\ldots,N)$ and
$E^\pm$, so we can get the dark-like solution for Eq. \eqref{eig11}
from the solution to Eq. \eqref{eig1} for the N-qubit Rabi model
just by making the replacement $f(\Delta_k,E^\pm)=0\rightarrow
f(\Delta_k,(E^\pm+M-1))=0$. To conclude, for a dark-like state of
the N-qubit Rabi model,we can get a corresponding dark-like state of
the N-qubit and M-photon Rabi model in the subspace labeled by
$\{i,\pm\}$, upon using the following relations
\begin{eqnarray}
E_{i}^\pm=E^\pm+i+M-1\\
g_{i,k}=\sqrt{\frac{(i+M)!}{i!}}g_k(k=1,2,\ldots,N)\\
f(\Delta_{i\pm,k},E_{i}^\pm-i)=f(\Delta_k,E^{\pm})=0.
\end{eqnarray}
As discussed above, the dark-like eigenstates of $H_{PQ}$ \eqref{gtq}
exist for arbitrary photon number $M$ in the whole qubit-photon
coupling regime with constant energy, even though generally the
model is only solvable under some constraints on the coupling
strength and photon number $M$.

For the two-qubit and two-photon Rabi model, there are six dark-like
states \begin{eqnarray}
|\psi\rangle_{0,+}=\frac{2(\Delta_1-\Delta_2)}{\sqrt{2}g}|0,\uparrow,\uparrow\rangle
-|2,\uparrow,\downarrow\rangle+|2,\downarrow,\uparrow\rangle,\label{dk210}\\
|\psi\rangle_{1,+}=\frac{2(\Delta_1-\Delta_2)}{\sqrt{6}g}|1,\uparrow,\uparrow\rangle
-|3,\uparrow,\downarrow\rangle+|3,\downarrow,\uparrow\rangle,\label{dk211}
\end{eqnarray}
with the conditions $g_1=g_2=g/2$, $\Delta_1+\Delta_2=2$ and
$E^+=2,3$ respectively, and
\begin{eqnarray}
|\psi\rangle_{0,-,a}=\left(\frac{2(\Delta_1+\Delta_2)}{\sqrt{2}g}|0,\uparrow,\downarrow\rangle+
 |2,\downarrow,\downarrow\rangle
-|2,\uparrow,\uparrow\rangle\right),\label{dk22}\\
|\psi\rangle_{1,-,a}=\left(\frac{2(\Delta_1+\Delta_2)}{\sqrt{6}g}|1,\uparrow,\downarrow\rangle+
 |3,\downarrow,\downarrow\rangle
-|3,\uparrow,\uparrow\rangle\right),\label{dk22}
\end{eqnarray}
with the conditions $g_1=g_2=g/2$, $\Delta_1-\Delta_2=2$ and
$E^-=2,3$ respectively, and
\begin{eqnarray}
 |\psi\rangle_{0,-,b}=\left(\frac{2(\Delta_1+\Delta_2)}{\sqrt{2}g}|0,\downarrow,\uparrow\rangle+
 |2,\downarrow,\downarrow\rangle
-|2,\uparrow,\uparrow\rangle\right),\label{dk23}\\
 |\psi\rangle_{1,-,b}=\left(\frac{2(\Delta_1+\Delta_2)}{\sqrt{6}g}|1,\downarrow,\uparrow\rangle+
 |3,\downarrow,\downarrow\rangle
-|3,\uparrow,\uparrow\rangle\right),\label{dk23}
\end{eqnarray}
with the conditions $g_1=g_2=g/2$, $\Delta_2-\Delta_1=2$ and
$E^-=2,3$ respectively.

Choosing $g_1=g_2=g/2$, $\Delta_1+\Delta_2=2$, the spectrum of the
two-qubit and two-photon Rabi model is shown in Figure 3, where the
dark-like states \eqref{dk210} and \eqref{dk211} correspond to the
horizontal line $E^+=2$ and $E^+=3$ respectively. These special
states exist in the whole coupling regime, while other eigenstates
exist only for $g<0.5$. Besides, they commonly exist even for
multi-qubit and M-photon ($M>2$) Rabi model. This can be tested by
numerical diagonalization: Although the eigenvalues usually will not
converge for $M=3$, with regard to dark-like states, the eigenvalue
always converge at $E=3$ with $i=0$.

\begin{figure}[htbp]
\center
\includegraphics[width=0.8\textwidth]{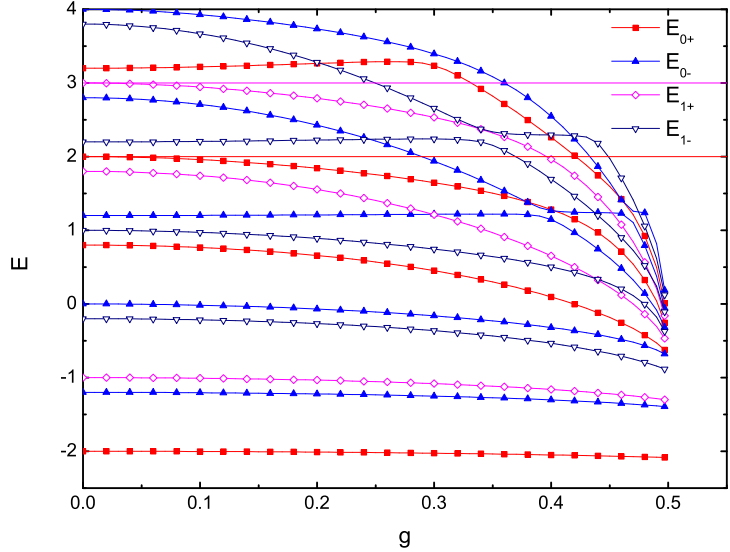}
\renewcommand\figurename{\textbf{Figure}}
\caption[2]{The numerical spectrum of two-qubit quantum Rabi model
with
$\Delta_1=1.6$,~$\Delta_2=0.4$,~$\omega=1$,~$g_1=g_2=g/2$,~$0\leq
g\leq0.5$. $E_{0+}$,~$E_{0-}$,~$E_{1+}$,~$E_{1-}$ are eigenvalues of
four invariant subspace labeled by $(i,\pm)$ respectively.
\label{figure3}}
\end{figure}

\section{Experimental considerations}\label{SecImplementation}
In the past few years there have appeared a series of proposals for the implementation of the quantum Rabi model in all its parameter regimes, via analog or digital-analog quantum simulations, in a variety of quantum platforms including trapped ions~\cite{Pedernales,Felicetti} and superconducting circuits~\cite{Ballester}.
Moreover, the multiqubit, single-photon Rabi model may be straightforwardly implemented in superconducting circuits via a digital-analog quantum simulator~\cite{MezzacapoRabi,LamataDicke}. Indeed, a set of superconducting qubits capacitively coupled with a coplanar microwave resonator naturally implement a Tavis-Cummings Hamiltonian. Via digital-analog techniques, one can combine this naturally-appearing interaction with local rotations, in order to reproduce the multiqubit Rabi model in all parameter regimes, and with arbitrary inhomogeneous couplings and qubit energies, with polynomial resources~\cite{MezzacapoRabi,LamataDicke}. Therefore, a quantum dynamics provided by the Hamiltonian in Eq.~(\ref{gq}) can be carried out in the lab with current technology. In order to probe the dark-like states of the multiqubit, single-photon Rabi model, one may proceed initializing the system in an eigenstate of an easy to initialize Hamiltonian, e.g., the purely qubit and bosonic mode free terms without mutual interaction, and adiabatically turn on the multiqubit Rabi coupling term, via a digitization of the adiabatic evolution, as in Ref.~\cite{Barends16}. In order to measure the energy, to check its constant character under parameter change, one may either apply the phase estimation algorithm, or measure term by term of the Hamiltonian, with standard superconducting circuit technology~\cite{MezzacapoRabi,LamataDicke}.

\section{Conclusions}\label{s4}
We have found dark-like states for multi-qubit and multi-photon Rabi
models, which exist in the whole coupling regime with constant
eigenenergy, with qubit and photon field still being coupled.
Besides, their photon numbers are bounded from above, distinctly
different from the one qubit case, because there are closed
subspaces in Fock space due to the interaction between multi-qubit
and photon field. Their existence conditions are simple, which does
not depend on qubit energy and coupling strength at the same time.
And they correspond to horizontal lines in the spectra, which means
for arbitrary coupling $g_i$, we always find one such state by
tuning other conditions. These dark-like states can also serve as
benchmarks for numerical techniques and as foundations for
perturbative treatments.

For the single-qubit and multi-photon Rabi model, the solution
exists only if the photon number $M\leq2$ and the coupling strength
is below a certain critical value. But multi-qubits make it
different. There exist dark-like eigenstates in the whole coupling
regime for arbitrary $M$ under certain conditions. This is due to
the closed subspace in the photon number representation brought
about by the multi-qubit, so just like the multi-photon J-C model,
the multi-photon Rabi model is diagonalizable in this special case.

Dark states can preserve
entanglement under dissipation, driving and dipole-dipole
interactions, so they can be used to store correlations. Dark-like states
have similar properties as dark states in the spectra, but their
properties under the influence of environment (dissipation,
dephasing, or the like) need to be explored. Whether this kind of dark-like
states has similar applications as dark-like states or has other
peculiarities is a very interesting problem to study.

\section*{Acknowledgements}
This work was supported by the National Natural Science Foundation
of China (Grants Nos 11535004, 11347112, 11204263, 11035001,
11404274, 10735010, 10975072, 11375086 and 11120101005), by the 973
National Major State Basic Research and Development of China (Grants
Nos 2010CB327803 and 2013CB834400), by CAS Knowledge Innovation
Project No. KJCX2-SW-N02, by Research Fund of Doctoral Point (RFDP)
Grant No. 20100091110028, by the Project Funded by the Priority
Academic Program Development of Jiangsu Higher Education
Institutions (PAPD), by the Scientific Research Fund of Hunan
Provincial Education Department (No. 12C0416), by the Program for
Changjiang Scholars and Innovative Research Team in University
(IRT13093), by Spanish MINECO/FEDER FIS2015-69983-P, UPV/EHU UFI
11/55, and Ram\'on y Cajal Grant RYC- 2012-11391.

%
%

\end{document}